\documentclass[a4paper]{article}

\usepackage[utf8]{inputenc}
\usepackage{amsthm}
\usepackage{amsmath}
\usepackage{amssymb}
\usepackage{latexsym}
\usepackage{enumitem}
\usepackage{xspace}
\usepackage{ifpdf}
\usepackage{algorithm}
\usepackage{algorithmic}
\usepackage{tikz}

\newtheorem{theorem}{Theorem}[section]
\newtheorem{lemma}[theorem]{Lemma}

\newtheorem{corollary}[theorem]{Corollary}

\newcommand{\cohesion}{\mathcal{C}}

\newcommand{\gcl}{{\sc Connected-Cohesive}\xspace}

\newcommand{\kclique}{{\sc Clique}\xspace}

\newcommand{\degree}{d}

\usepackage{RR}

\RRetitle{ Maximizing the Cohesion is NP-hard }
\RRtitle{ Maximiser la Cohésion est NP-dur }
\RRdomaine{3}
\RRtheme{R\'eseaux et t\'el\'ecommunications}
\RRauthor{
  Adrien Friggeri
  \and
  Eric Fleury
}
\URRhoneAlpes
\RRequipe{DNET}
\RRdate{8 September 2011}
\RRdater{3 October 2011}
\RRversion{2}

\RRmotcle {r\'eseaux sociaux, r\'eseaux complexes, co\'esion, np-complet, compl\'exit\'e}
\RRkeyword {social networks, complex networks, cohesion, np-complete, complexity}
\RRabstract{
We show that the problem of finding a set with maximum cohesion in an
undirected network is \textbf{NP}-hard.
}

\RRresume{ 
Nous montrons que le problème de trouver un ensemble de cohésion maximum dans
un graphe non orient\'e est \textbf{NP}-dur.
}

\begin{document}
\RRNo{7734}

\makeRR

\section*{Introduction}

In \cite{FRIGGERI:2011:INRIA-00619092:1}, we have introduced a new metric
called the \emph{cohesion} which rates the community-ness of a group of people
in a social network from a sociological point of view. Through a large scale
experiment on Facebook, we have established that the cohesion is highly
correlated to the subjective user perception of the communities. In this
article, we show that finding a set of vertices with maximum cohesion is
\textbf{NP}-hard.

\section*{Notations}

Let $G=(V,E)$ be a graph with vertex set $V$ and edge set $E$ of size
$n=|V|\geq 4$. For all vertices $u\in V$, we write $\degree_G(u)$ the degree
of $u$, or more simply $\degree(u)$\footnote{Here, as elsewhere, we drop the
index referring to the underlying graph if the reference is clear.}. A
\emph{triangle} in $G$ is a triplet of pairwise connected vertices.

For all sets of vertices $S\subseteq V$, let $G[S]=(S,E_S)$ be the subgraph
induced by $S$ on $G$. We write $m(S)=|E_S|$ the number of edges in $G[S]$,
and $i(S)=|\{(u,v,w)\in S^3 : (uv,vw,uw) \in E^3\}|$ the number of
triangles in $G[S]$. We define $o(S) = |\{(u,v,w), (u,v)\in S^2, w\in
V\setminus S : (uv,vw,uw)\in E^3\}|$, the number of \emph{outbound}
triangles of $S$, that is: triangles in $G$ which have exactly two vertices in
$S$.

Moreover, for all $(u,v)$ in $E$, let $\triangle(uv)=|\{w\in V :
(uw,vw)\in E^2\}|$ be the number of triangles the edge $uv$ belongs
to in $G$.

Finally, we recall the definition of the cohesion of a set $S$ in $G$:
$$\cohesion(S)=\frac{i(S)^2}{{|S| \choose 3}(i(S)+o(S))}$$

An example is given on Figure~\ref{fig:cohesion_example}. The cohesion of a
given set $S$ in $G$ can naively be computed in $\mathcal{O}(n^3)$ by listing
all triangles in $G$ and counting those inside and outbound to $S$.

\begin{figure}[htb]
  \centering
  \begin{tikzpicture}[scale=1]
    
    \filldraw [rounded corners=12,color=black!20,fill=black!8] (-0.8,-0.5) rectangle (2.5, 2.5);
    \draw (-0.5,1) node {$S$};
    \draw [line width=0.3mm,color=black!65] (0,0) -- (2,2) -- (0,2) -- (0,0) -- (2,0) -- (3.6,1) -- (2,2) -- (2,0);
    \draw [line width=0.2mm,color=black!65,fill=black!50] (0,0) circle (0.3);
    \draw [line width=0.2mm,color=black!65,fill=black!50] (0,2) circle (0.3);
    \draw [line width=0.2mm,color=black!65,fill=black!50] (2,0) circle (0.3);
    \draw [line width=0.2mm,color=black!65,fill=black!50] (2,2) circle (0.3);
    \draw [line width=0.2mm,color=black!65,fill=black!50] (3.6,1) circle (0.3);
    
  \end{tikzpicture}
  \caption{In this example, $i(S)=2$ and $o(S)=1$, thus $\cohesion(S)=\frac{1}{6}$}
  \label{fig:cohesion_example}
\end{figure}

In this article we examine the problem of finding a set of vertices $S\subseteq V$
of maximum cohesion, i.e. for all subset $S^\prime\subseteq V$,
$\cohesion(S^\prime)\leq\cohesion(S)$.

\subsection*{Outline}

We now proceed to prove that finding a set of vertices with maximum cohesion in
$G$ is \textbf{NP}-hard. We will first show in Section~\ref{sec:maxconnected}
that this problem is equivalent to that of finding a connected set of vertices
with maximum cohesion in $G$. The decision problem associated to the latter is
\gcl.

Then, we shall prove that \gcl is \textbf{NP}-complete by reducing \kclique
(problem GT19 in \cite{Garey1979}). From there we deduce that the optimization
problem of finding a set of vertices with maximum cohesion is
\textbf{NP}-hard.

\paragraph{Problems}
\begin{enumerate}
  \item \gcl:
    \label{pb:gcl}
    \begin{description}[noitemsep=nosep,style=multiline,leftmargin=1.8cm]
      \item[Input]    A graph $G=(V,E)$, $\lambda\in\mathbb{Q}$, $\lambda\in[0,1]$
      \item[Question] Is there a subset connected $S$ of $V$ such that $\cohesion(S) \geq \lambda$?
    \end{description}
  \item \kclique:
    \label{pb:kclique}
    \begin{description}[noitemsep=nosep,style=multiline,leftmargin=1.8cm]
      \item[Input]    A graph $G=(V,E)$, $k\in\mathbb{N}, k\leq |V|$
      \item[Question] Is there a subset $S$ of $V$ such that $|S|=k$ and the subgraph induced by $S$ is a clique?
    \end{description}
\end{enumerate}

\section{A maximum cohesive group is connected}
\label{sec:maxconnected}

In order to prove that a set of vertices with maximum cohesion in a given network
is connected, we need the following lemma:

\begin{lemma}
  \label{thm:lemconnected}
  Let $S_1\subseteq V$ and $S_2\subseteq V$ be two disconnected sets of vertices
  $((S_1\times S_2) \cap E = \emptyset)$. If $\cohesion(S_1) \leq
  \cohesion(S_1\cup S_2)$ then $\cohesion(S_2) >
  \cohesion(S_1\cup S_2)$.
  
  \begin{proof}
    Suppose $\cohesion(S_1) \leq \cohesion(S_1\cup S_2)$ and
    $\cohesion(S_2) \leq \cohesion(S_1\cup S_2)$.
    Given that $S_1$ and $S_2$ are disconnected, $i(S_1\cup
    S_2)=i(S_1)+i(S_2)$ and $o(S_1\cup S_2)=o(S_1)+o(S_2)$. We can
    then write:
    \begin{align}
      \frac{i(S_1)^2}{{|S_1| \choose 3}} &\leq (i(S_1)+o(S_1))\cohesion(S_1\cup S_2) \\
      \frac{i(S_2)^2}{{|S_2| \choose 3}} &\leq (i(S_2)+o(S_2))\cohesion(S_1\cup S_2)
    \end{align}
    By summing (1) and (2), we obtain:
    \begin{align*}
      \frac{i(S_1)^2}{{|S_1| \choose 3}} + \frac{i(S_2)^2}{{|S_2| \choose 3}}
        &\leq (i(S_1)+o(S_1)+i(S_2)+o(S_2))\cohesion(S_1\cup S_2)\\
        &\leq (i(S_1\cup S_2)+o(S_1\cup S_2))\cohesion(S_1\cup S_2)\\
        &\leq \frac{(i(S_1)+i(S_2))^2}{{{|S_1| + |S_2|} \choose 3}}
    \end{align*}
    Furthermore, given that $|S_1|,|S_2|>1$,
    $${|S_1| \choose 3} + {|S_2| \choose 3} < {|S_1|+|S_2| \choose 3}$$
    We then have:
    \begin{align*}
      \frac{i(S_1)^2}{{|S_1| \choose 3}} + \frac{i(S_2)^2}{{|S_2| \choose 3}} 
        &< \frac{(i(S_1)+i(S_2))^2}{{|S_1| \choose 3} + {|S_2| \choose 3}}
    \end{align*}
    Which simplifies to:
    \[
      \left({|S_2| \choose 3}i(S_1) - {|S_1| \choose 3}i(S_2)\right)^2 < 0
    \]
    Hence the contradiction. Therefore, for all $S_1, S_2\subseteq V$, disconnected:
    $$\cohesion(S_1) \leq \cohesion(S_1\cup S_2) \Rightarrow \cohesion(S_2) > \cohesion(S_1\cup S_2)\qedhere$$
  \end{proof}
\end{lemma}

\begin{theorem}
  \label{thm:connected}
  Let $S$ be the set of vertices of $G$ with the highest cohesion, $S$ is connected.
  
  \begin{proof}
    Suppose $S$ is not connected, then their exist two disconnect subsets
    $S_1, S_2 \subseteq S$ such that $S=S_1\cup S_2$. Given that $S$ has
    maximum cohesion, we have $\cohesion(S)\geq\cohesion(S_1)$. Thus per
    Lemma~\ref{thm:lemconnected}: $\cohesion(S)<\cohesion(S_2)$ and $S$ does
    not have the highest cohesion, hence the contradiction.
  \end{proof}
\end{theorem}

\begin{corollary}
  Per Theorem~\ref{thm:connected}, the problem of searching for a set of vertices
  with maximum cohesion is strictly equivalent to that of searching a set of
  connected vertices with maximum cohesion.
\end{corollary}

\section{\gcl is \textbf{NP}-complete}
\label{sec:gcldnp}

First note that given a set $S$ of vertices of $G$, it is possible to verify
that $S$ is a solution of \gcl by computing its cohesion, its size, its
connectivity and the minimum degree of its vertices, all in polynomial time.
Therefore \gcl is in \textbf{NP}.

\newcommand{\K}{\ensuremath{2{n \choose 3}^4}}
\newcommand{\target}{\ensuremath{\frac{{k \choose 3}}{{k \choose 3}+{k \choose 2}(n-k)}}}
\begin{algorithm}[htb]
  \caption{Transforms an instance of \kclique in an instance of \gcl}
  \label{alg:spiking}
  \begin{algorithmic}[1]
    \REQUIRE $G=(V,E), k\in\mathbb{N}$
    \STATE $W := \emptyset$
    \STATE $E':=E$
    \FOR{$uv\in V^2 \setminus E$}
      \STATE let $K$ be a clique of size \K
      \STATE $W\gets W\cup K$
      \STATE $E'\gets E' \cup \{uv\} \cup (\{u,v\}\times K)$
    \ENDFOR
    \RETURN $G'=(V\cup W,E'), \lambda=\target$
  \end{algorithmic}
\end{algorithm}

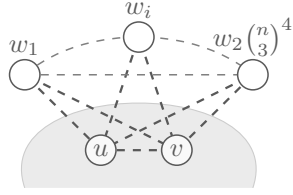
\begin{figure}[htb]
  \centering
  \begin{tikzpicture}[scale=1]
    
    \filldraw [color=black!20,fill=black!8] (-1.5,0.5) .. controls (-2, 2) and (2,2) .. (1.5,0.5);
    \draw [line width=0.3mm,color=black!65,style=dashed] (-0.5,1) -- (0.5,1);
    \draw [line width=0.3mm,color=black!65,style=dashed] (-0.5,1) -- (-1.5, 2) -- (0.5,1) -- (0,2.5) -- (-0.5,1) -- (1.5, 2) -- (0.5,1) ;
    \draw [line width=0.2mm,color=black!50,style=dashed] (-1.5,2) .. controls (-1.5, 2) and (-1,2.5) .. (0,2.5) .. controls (1,2.5) and (1.5, 2) ..  (1.5,2);
    \draw [line width=0.2mm,color=black!50,style=dashed] (-1.5,2) -- (1.5,2);
    \draw [line width=0.2mm,color=black!65,fill=gray!5] (-0.5,1) circle (0.2) node {$u$};
    \draw [line width=0.2mm,color=black!65,fill=gray!5] (0.5,1) circle (0.2) node {$v$};
    \draw [line width=0.2mm,color=black!65,fill=white] (-1.5,2) circle (0.2) node [label=92:$w_1$] {};
    \draw [line width=0.2mm,color=black!65,fill=white] (0,2.5) circle (0.2) node [label=90:$w_i$] {};
    \draw [line width=0.2mm,color=black!65,fill=white] (1.5,2) circle (0.2) node [label=88:$w_\K$] {};
    
  \end{tikzpicture}
  \caption{Illustration of Algorithm~\ref{alg:spiking}. At this step, wejoin
  $u$ and $v$, add a clique of size $\K$ to the network, and join $u$ and $v$
  to all vertices in the added clique.}
  \label{fig:spiking}
\end{figure}

Let us now reduce \kclique to \gcl. Let $(G=(V,E), k\in\mathbb{N})$ be an
instance of \kclique\footnote{We consider here that $|G|>2$ and $k>2$,
although this is not exactly \kclique, this problem is clearly
\textbf{NP}-complete, given that the complexity of \kclique does not arise
from those small values.}. We can assume that $G$ is connected (if not, we use
the following reasoning separately on each connected component of $G$). We
construct an instance $(G^\prime=(V^\prime, E^\prime), \lambda)$ of \gcl by
adding an edge between all non connected vertices $u$ and $v$ in $G$ and then
linking those two vertices to all vertices in a clique of size \K which we add
to the network, as described in Algorithm~\ref{alg:spiking} and illustrated by
Figure~\ref{fig:spiking}.

\begin{theorem}
  \label{thm:clique}
  There exist a clique of size $k$ in $G$ \emph{iff} there exist a connected
  group of vertices of $G^\prime$ with cohesion $\lambda\geq\target$.
  
  \begin{proof}
    Let $K\subseteq V$, be a clique of size $|K|=k$ in $G$. Given that no node
    or edges are deleted when constructing $G^\prime$, $G$ is a subgraph of
    $G^\prime$ and thus $K$ is a clique in $G^\prime$ and $i_{G^\prime}(K) =
    {k \choose 3}$. 
    
    Moreover, by construction, $G^\prime[V]$ is a clique and for all $u$ un
    $K$, the neighbors of $u$ are also in $V$. Therefore, each edge in $K$
    forms one triangle with each vertex in $V\setminus K$, which leads to
    $o_{G^\prime}(K) = {k \choose 2}(n-k)$. Finally, this gives a cohesion:
    $$\cohesion_{G^\prime}(K)=\target$$
    
    Conversely, let $S \subseteq V^\prime$ be a connected set of vertices such
    that $\cohesion_{G^\prime}(S) \geq \target$. We will show that $S$ is a
    clique of size larger than $k$ and that $S\subseteq V$. First note that
    $|S|\geq 3$, because by definition, if $|S|<3$,
    $\cohesion_{G^\prime}(S)=0$ which would lead to a contradiction.
    
    First, suppose that $S$ is not a clique in $G$, then let us distinguish
    two cases:
    \begin{enumerate}
      \item If $S\subseteq V$ and $S$ is not a clique, then $S$ contains two
      vertices $u,v\in V^2$ such that $uv\not\in E$.
      \item If $S\not\subseteq V$, then $\exists u\in S\setminus V$, and $S$
      being connected, there exist $v\in V^\prime$ such that $uv\not\in E$.
    \end{enumerate}
    Therefore, if $S$ is not a clique in $G$, it contains an edge $uv\not\in
    E$ and by construction, this edge belongs to at least $\K$ triangles,
    which leads to:
    \begin{align*}
      i_{G^\prime}(S)+o_{G^\prime}(S) &\geq K \\
      \cohesion_{G^\prime}(S) &\leq\frac{i_{G^\prime}(S)^2}{2{|S| \choose 3}{n \choose 3}^4}  \\
                              &\leq\frac{1}{2{n\choose 3}^2}\\
                              &< \target
    \end{align*}
    Hence the contradiction, therefore $S$ must be a clique in $G$. From there
    it comes that:
    $$\cohesion_{G^\prime}(S)=\frac{{k^\prime \choose 3}}{{k^\prime \choose 3}+{k^\prime \choose 2}(n-k^\prime)}$$
    where $k^\prime = |S|$. Therefore:
    \begin{align*}
      \cohesion_{G^\prime}(S)\geq\target &\Leftrightarrow \frac{{k^\prime \choose 2}(n-k^\prime)}{{k^\prime \choose 3}}\leq\frac{{k \choose 2}(n-k)}{{k \choose 3}}\\
                                         &\Leftrightarrow \frac{n-k^\prime}{k^\prime-3} \leq \frac{n-k}{k-3}\\
                                         &\Leftrightarrow k^\prime \geq k
    \end{align*}
    Therefore, we can now conclude that if there exist a connected set $S$ in
    $G^\prime$ with cohesion $\cohesion_{G^\prime}(S)\geq\target$, then $S$ is
    a clique of size at least $k$ in $G$, and thus there exist a clique
    $K\subseteq S$ of size $k$ in $G$.
  \end{proof}

\end{theorem}

\begin{theorem}
  \gcl is \textbf{NP}-complete.
  
  \begin{proof}
    Per Theorem~\ref{thm:clique}, there exist a clique of size $k$ in $G$
    \emph{iff} there exist a connected subset of vertices of $G^\prime$ of
    cohesion $\lambda\geq\target$ and the transformation from $G,k$ to
    $G^\prime,\lambda$ runs in polynomial time. Thus \kclique is reducible
    to \gcl and \gcl is \textbf{NP}-hard.
    
     Given that \gcl is in \textbf{NP}, the problem is thus
    \textbf{NP}-complete.
  \end{proof}
\end{theorem}

\section{Conclusion}

The associated decision problem being \textbf{NP}-complete, the problem of
finding a set of vertices with maximum cohesion is
\textbf{NP}-hard\footnote{Note that the problem of finding a set of vertices
of maximum cohesion containing a set of predefined vertices is also
\textbf{NP}-hard, by an immediate reduction}.


\bibliographystyle{plain}
\bibliography{biblio}

\begin{thebibliography}{1}

\bibitem{FRIGGERI:2011:INRIA-00619092:1}
Adrien Friggeri, Guillaume Chelius, and Eric Fleury.
\newblock {Triangles to Capture Social Cohesion}.
\newblock In {\em {Third IEEE International Conference on Social Computing}},
  Cambridge, United States, September 2011.

\bibitem{Garey1979}
M.R. Garey and D.S. Johnson.
\newblock {\em Computers and Intractability: A Guide to the Theory of
  NP-Completeness}.
\newblock W.H. Freeman, San Francisco, 1979.

\end{thebibliography}
\end{document}